\documentclass[showpacs,preprintnumbers,amsmath,amssymb,twocolumn,superscriptaddress,prb]{revtex4}
\usepackage{times}
\usepackage{graphicx}
\usepackage{amsfonts}
\usepackage{amsmath, amsthm, amssymb}
\usepackage{dsfont}

\newcommand{\ml}{\boldsymbol{M}_\text{L}}
\newcommand{\mr}{\boldsymbol{M}_\text{R}}

\newcommand{\e}[1]{\mathrm{e}^{#1}}

\newcommand{\eg}{\textit{e.g. }}

\def\i{\mathrm{i}}

\begin{document}
\title[Spin-Transfer Torque and Magnetoresistance in Superconducting Spin-Valves]{Spin-Transfer Torque and 
Magnetoresistance in Superconducting Spin-Valves}
\author{Jacob Linder}
\affiliation{Department of Physics, Norwegian University of
Science and Technology, N-7491 Trondheim, Norway}
\author{Takehito Yokoyama}
\affiliation{Department of Applied Physics, Nagoya University, Nagoya, 464-8603, Japan}
\author{Asle Sudb{\o}}
\affiliation{Department of Physics, Norwegian University of
Science and Technology, N-7491 Trondheim, Norway}

\date{Received \today}
\begin{abstract}
We study the spin-transfer torque and magnetoresistance of a ferromagnet$\mid$superconductor$\mid$ferromagnet spin-valve, 
allowing for an arbitrary magnetization misorientation and treating both $s$-wave and $d$-wave symmetries of the 
superconductor. We take fully into account Andreev reflection and also the spin-triplet correlations that are 
generated when the magnetizations are non-collinear. It is found that the torque and magnetoresistance are both 
strongly enhanced when topological zero-energy states are present at the interfaces, which is the case for 
$d$-wave superconductors with a crystallographic orientation of [110] relative to the interface ($d_{xy}$-wave symmetry). 
Moreover, we find that the magnetoresistance displays a strong oscillatory and non-monotonous behavior as a function 
of $d_S/\xi$ where $d_S$ and $\xi$ are the interlayer width  of the superconducting region and the superconducting 
coherence length, respectively. This feature is also attributed to the crossover from layers of size $d_S\sim 2\xi$ 
to layers of size $d_S\gg 2\xi$, where the contribution to transport from zero-energy states gradually vanishes.
\end{abstract}

\maketitle

\section{Introduction}
The magnetization dynamics induced by electric currents is a topic presently benefitting from great interest 
\cite{zutic_rmp_04, tserkovnyak_rmp_05, brataas_physrep_06,Maekawa,Tatara, ralph_jmmm_08}. Typically, the dynamics 
is studied in spin-valve or magnetic tunnel junction setups consisting of two or more ferromagnetic layers with 
misaligned magnetization orientations. These layers routinely consist of segments with free and fixed magnetization 
directions separated by non-magnetic spacers in order to overcome the exchange coupling between them. While the
 magnetoresistance (MR) of such structures has been the focus of most investigations so far, there are other 
 intriguing phenomena occuring in this type of systems that harbor a great potential for applications. One 
 of these phenomena is the so-called \textit{spin-transfer torque} \cite{ralph_jmmm_08, sankey_nphys_08}. The 
 basic concept of spin-transfer torque is illustrated by considering a spin-current incident on a ferromagnetic 
 layer with a magnetization that is misaligned compared to the polarization of the spin-current. Upon entering 
 the ferromagnetic region, a portion of the incoming spin-current will in general be absorbed by the 
 ferromagnetic order parameter, causing the magnetization to precess. This suggests that if the incoming 
 spin-current is large enough, it may actually switch the magnetization direction of the second ferromagnet 
 \cite{slonczewski_jmmm_96}, an effect which has been observed experimentally \cite{myers_science_99, katine_prl_00}. 
 The spin-transfer torque then arises due to the non-conservation of the spin-current.
\par
While the conventional spin-valve structure of type ferromagnet$\mid$normal metal$\mid$ferromagnet (F$\mid$N$\mid$F) 
has been the subject of much investigation, magnetization dynamics and spin-transfer torques in the superconducting 
(S) analogue junction, F$\mid$S$\mid$F junction, has received less attention so far. Only a few works have studied 
effects related to the spin-transfer torque in such heterostructures 
\cite{tserkovnyak_prb_02, waintal_prb_01, waintal_prb_02, lofwander_prl_05,Zhao}. In particular, the authors of 
Ref. \cite{tserkovnyak_prb_02} calculated the spin-transfer torque in an F$\mid$S$\mid$F spin-valve for an arbitrary 
misorientation angle between the ferromagnets when disregarding the effect of Andreev-reflection. However, 
Andreev-reflection is in general present, and should be expected to heavily influence the resulting spin-transfer 
torque. Also, equal-spin pairing $(S_z=\pm1)$ superconducting proximity amplitudes are induced in the system 
whenever the magnetizations are non-collinear.\cite{bergeretrmp,buzdinrmp,Braude,Houzet} Therefore, it remains 
to be clarified precisely how the superconducting correlations affect the spin-transfer torque in F$\mid$S$\mid$F 
spin-valves. In addition, to the best of our knowledge there exists no study of the spin-transfer torque in 
F$\mid$S$\mid$F systems where the superconductor has an unconventional pairing symmetry, such as $d$-wave, 
which would be relevant for spin-valve setups with high-$T_c$ cuprate superconductors.

\begin{figure}[b!]
\centering
\resizebox{0.5\textwidth}{!}{
\includegraphics{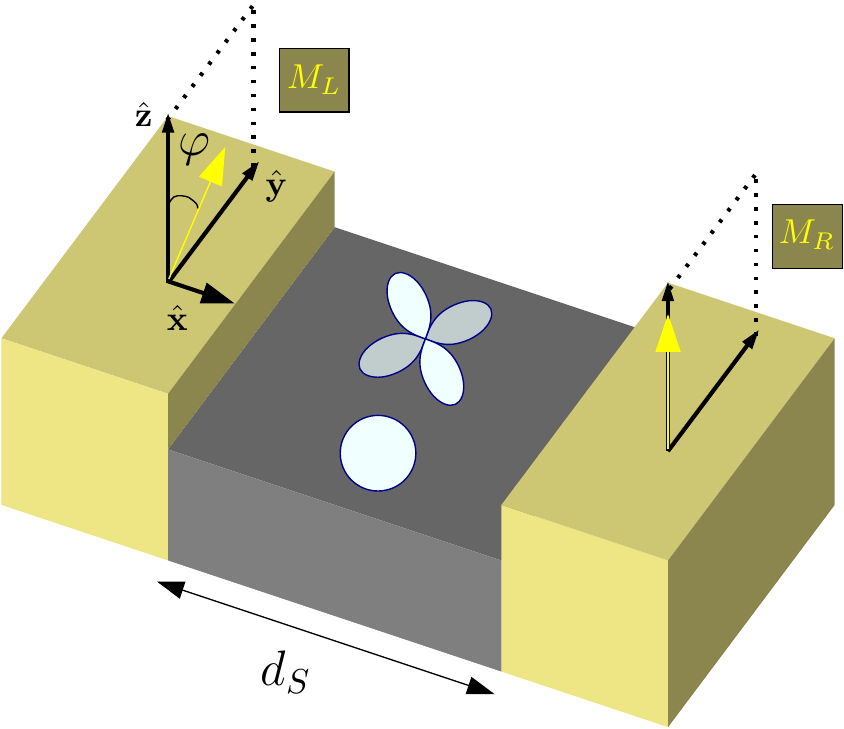}}
\caption{(Color online) A spin-valve setup with misaligned magnetizations in the ferromagnetic regions. We consider 
both an $s$-wave and $d$-wave symmetry in the superconducting region.}
\label{fig:model}
\end{figure}

In this paper, we calculate the spin-transfer torque in a F$\mid$S$\mid$F spin-valve (see Fig. \ref{fig:model}) and 
fully take into account both Andreev reflection and triplet correlations that are present in the system. As mentioned, 
we will allow for an anisotropic superconducting order parameter such as $d$-wave in order to investigate how this 
influences the spin-transfer torque. Moreover, we investigate the MR effect of the spin-valve, with particular 
emphasis on the influence of topological zero-energy states which may be present when the superconductor has a 
$d$-wave symmetry\cite{Hu,Tanaka,Ryu}. Our main finding is that both the spin-transfer torque and MR are enhanced 
considerably when topological zero-energy states are present at the interfaces, as is relevant for the $d_{xy}$-wave 
case. Also, the MR exhibits a non-monotonous behavior as a function of the width $d_S$ of the superconductor, an 
effect which we will show pertains to the existence of the zero-energy states.
\par
We organize this work as follows. In Sec. \ref{sec:theory}, we introduce the theoretical framework used to obtain 
our results, which are presented in Sec. \ref{sec:results}. We discuss our results in Sec. \ref{sec:discussion}, 
and give concluding remarks in Sec. \ref{sec:summary}.

\section{Theory}\label{sec:theory}

Our model is shown in Fig. \ref{fig:model}. We consider transport in the $x-y$ plane, while the magnetizations of the F 
layers are allowed to rotate in the $y-z$ plane. This is the most relevant setup experimentally. We are considering a 
spin-valve setup where an electrical current flows from ferromagnetic region F$_\text{L}$ ($x<0$) where it becomes 
spin-polarized. The current then passes through a superconducting region ($0<x<d_S$) and into a second ferromagnetic 
region F$_\text{R}$ ($x>d_S$), on which it exerts a spin-torque. The Bogolioubov-de Gennes (BdG) equations relevant 
for our system are formally derived by means of diagonalizing the Hamiltonian. In their full form, the BdG equations 
read:
\begin{align}
\begin{pmatrix}
\hat{H}_0  & \Delta(x)\i\hat{\sigma}_2\\
[\Delta(x)]^*\i\hat{\sigma}_2 & -\hat{H}_0^\text{T} \\
\end{pmatrix} \psi
= \varepsilon\psi,
\end{align}
where we have defined the normal-state Hamiltonian $\hat{H}_0$ and wavefunction $\psi$:
\begin{align}
\notag\\
\hat{H}_0 = \Big[-\frac{\nabla^2}{2m(x)} - \mu(x)\Big] \hat{1} - \boldsymbol{h}(x)\cdot\hat{\boldsymbol{\sigma}},\;
\psi = \begin{pmatrix}
u_\uparrow\\
u_\downarrow\\
v_\uparrow\\
v_\downarrow\\
\end{pmatrix}
\end{align}
The superconducting order parameter is non-zero only for $0<x<d_S$ and the magnetic exchange field is given by $\boldsymbol{h}(x) =
 \boldsymbol{h}_\text{L}$ for $x<0$ and $\boldsymbol{h}(x) =  \boldsymbol{h}_\text{R} \parallel \hat{\boldsymbol{z}}$ for $x>d_S$. 
 Similar considerations apply to the chemical potential $\mu$ and the effective mass $m$. We denote a $2\times2$ matrix with 
 $\hat{\ldots}$, and the superscript 'T' denotes the matrix transpose. The exchange field is roughly related to the magnetization 
 $\boldsymbol{M}$ as $\boldsymbol{M} \simeq -\mu_BN_0\boldsymbol{h}$ in the quasiclassical approximation. The elements 
 $u_\sigma$ and $v_\sigma$ of the wavefunction may be interpreted as electron- and hole-like components with spin $\sigma$, 
 respectively. We have used units $\hbar$=$c$=1 and from now on use a real gauge for the superconducting order parameter.
\par
We will now use an extended Blonder-Tinkham-Klapwijk (BTK)-formalism \cite{btk} to construct the wavefunctions which account 
for the transport properties of the spin-valve, using results from Ref. \cite{linder_prb_07} to account for the misalignment 
angle between the magnetizations. A similar approach was used in Ref. \cite{xiao_prb_08} to investigate the spin-transfer 
torque in a conventional F$\mid$N$\mid$F spin-valve. For an incident plane-wave from the F$_\text{L}$ region with spin 
$\sigma$ with respect to the magnetization of  F$_\text{L}$  at an angle of incidence $\theta$ measured from normal to 
the interface, we obtain the following wavefunction in the left ferromagnetic region:
\begin{widetext}
\begin{align}\label{eq:wave}
\psi_\text{F$_\text{L}$} &= \begin{pmatrix}
\cos(\varphi/2)(s_\uparrow \e{\i k_e^\uparrow x} + r_e^\uparrow\e{-\i k_e^\uparrow x}) +\i \sin(\varphi/2)(s_\downarrow\e{\i k_e^\downarrow x} + r_e^\downarrow \e{-\i k_e^\downarrow x}) \\
\i\sin(\varphi/2)(s_\uparrow \e{\i k_e^\uparrow x} + r_e^\uparrow\e{-\i k_e^\uparrow x}) +\cos(\varphi/2)(s_\downarrow\e{\i k_e^\downarrow x} + r_e^\downarrow \e{-\i k_e^\downarrow x}) \\
r_h^\uparrow\cos(\varphi/2) \e{\i k_h^\uparrow x} - r_h^\downarrow \i\sin(\varphi/2)\e{\i k_h^\downarrow x}\\
-r_h^\uparrow\i\sin(\varphi/2) \e{\i k_h^\uparrow x} + r_h^\downarrow \cos(\varphi/2)\e{\i k_h^\downarrow x}\\
\end{pmatrix}.
\end{align}
\end{widetext}
Here, $\varphi$ is the misalignment between magnetizations of F layers. 
In the superconductor, one may write:
\begin{align}
\Psi_\text{S} &= \sum_\pm \begin{pmatrix}
a_\pm u_\pm \e{\pm\i q_{e}x} + d_\pm v_\pm \gamma_\pm \e{\pm\i q_{h}x} \\
b_\pm u_\pm \e{\pm\i q_{e}x}   - c_\pm v_\pm\gamma_\pm \e{\pm\i q_{h}x} \\
-b_\pm v_\pm \gamma_\pm^*  \e{\pm\i q_{e}x}  + c_\pm u_\pm  \e{\pm\i q_{h}x}\\
a_\pm v_\pm \gamma_\pm^* \e{\pm\i q_{e}x}  + d_\pm u_\pm \e{\pm\i q_{h}x}\\
\end{pmatrix},
\end{align}
whereas the wavefunction in the right ferromagnet reads:
\begin{align}
\psi_\text{F$_\text{R}$} =\begin{pmatrix}
t_e^\uparrow \e{\i k_e^\uparrow x}\\
t_e^\downarrow \e{\i k_e^\downarrow x}\\
t_h^\uparrow \e{-\i k_h^\uparrow x}\\
t_h^\downarrow \e{-\i k_h^\downarrow x}\\
\end{pmatrix}
\end{align}
Above, $\{s_\uparrow=1, s_\downarrow=0\}$ for $\sigma=\uparrow$ and $\{s_\uparrow=0, s_\downarrow=1\}$ for $\sigma=\downarrow$. In 
the superconducting regions, the wavefunctions are superpositions of left- and right-going electron-like and hole-like 
quasiparticles, with wave-vectors
\begin{align}
q_{e[h]} &=  \sqrt{2m_S(\mu_S +[-] \sqrt{\varepsilon^2-|\Delta(\theta_S)|^2}) - k_\perp^2}.
\end{align}
The momentum perpendicular to the interface is conserved, and given as 
\begin{align}
k_\perp = \sqrt{2m_F(\mu_F + \sigma h + \varepsilon)}\sin\theta
\end{align}
for incoming particles with spin $\sigma$. The wavevectors in the ferromagnetic region are:
\begin{align}\label{eq:vectork}
k_{e[h]}^\sigma &= \sqrt{2m_F(\mu_F + \sigma h +[-] \varepsilon) - k_\perp^2}.
\end{align}
Note that we are distinguishing the effective electron mass and the Fermi level in the ferromagnetic and superconducting region, in 
order to employ realistic values for these parameters in what follows. The wavevectors $q_{e[h]}$ in the superconducting region are 
written down under the quite general assumption that the gap satisfies $|\Delta(k_x,k_y)| = |\Delta(-k_x,k_y)|$, which is valid 
for both $s$-wave, $d_{xy}$-wave, and $d_{x^2-y^2}$-wave symmetries. The superconducting coherence factors are therefore given by
\begin{align}
u_\pm [v_\pm] = \frac{1}{\sqrt{2}}\Big[1 + [-] \frac{\sqrt{\varepsilon^2 - |\Delta(\theta_S)|^2}}{\varepsilon}\Big]^{1/2},
\end{align}
where the angle of propagation in the superconducting region is given by
\begin{align}
\theta_S = \text{asin}(k_\perp/\sqrt{2m_S\mu_S}).
\end{align}
The transport properties of the spin-valve are expected to be sensitive to the internal phase of the superconducting order parameter, which varies with $\theta$ in the $d$-wave case. To capture this effect, we have introduced the phase-sensitive factors 
\begin{align}
\gamma_\pm = \Delta(\theta_S^\pm)/|\Delta(\theta_S^\pm)|,\; \theta_S^+ = \theta_S,\; \theta_S^- = \pi-\theta_S.
\end{align}
For the $s$-wave symmetry we set $\Delta(\theta)=\Delta_0$, while for the $d$-wave case we set $\Delta(\theta) = \Delta_0\cos[2(\theta-\alpha)]$ with $\alpha=0$ for $d_{x^2-y^2}$-wave and $\alpha=\pi/4$ for $d_{xy}$-wave pairing.
\par
The presence of a ferromagnetic material may lead to a spin-dependent barrier potential at the interface. In general, there 
may also be spin-flip scattering processes at the interface due to magnetic inhomogeneities or misaligned moments in the 
interface region \cite{eschrig_naturephysics_08}. Such spin-flip scattering processes may induce equal-spin superconducting 
spin-triplet correlations. Since this type of correlations appear anyway due to the possibility of a misalignment between 
the magnetizations in the two F layers, we will not include such spin-flip processes at the interface. Quantitatively, the influence of
interfacial spin-flip processes compared to non-collinear magnetizations may be different.
Qualitatively, the effects brought about by the two should however remain the same. 
Here, we will focus 
primarily on the effect of the misalignment $\varphi$ between the F layers, and we therefore do not write any explicit 
spin-dependence of the barrier potential. Note that the scattered particles nevertheless experience spin-dependent 
phase-shifts due to the exchange potential in the F layers. \cite{Brataas}
\par
The scattering coefficients in Eq. (\ref{eq:wave}) are determined by using appropriate boundary conditions. Assuming 
a barrier potential of $V_0$ at each of the F$\mid$S interfaces, we may define a dimensionless measure $Z$ of the 
transparency:
\begin{align}
Z = 2m_FV_0/\sqrt{2m_F\mu_F}.
\end{align}
The boundary conditions now read:
\begin{align}\label{eq:bc}
\psi_\text{F$_\text{L}$} = \Psi_\text{S},\; \partial_x(\Psi_\text{S} - \psi_\text{F$_\text{L}$}) = 2m_FV_0\psi_\text{F$_\text{L}$},\notag\\
\psi_\text{F$_\text{R}$} = \Psi_\text{S},\; \partial_x(\psi_\text{F$_\text{R}$} - \Psi_\text{S}) = 2m_FV_0\psi_\text{F$_\text{R}$}.
\end{align}
The scattering coefficients are obtained numerically by setting up a system of equations
\begin{align}
\hat{\mathcal{A}}\boldsymbol{x} = \boldsymbol{b},
\end{align}
where the matrix $\hat{\mathcal{A}}$ and the vector $\boldsymbol{b}$ are constructed by inserting Eqs. (\ref{eq:wave}), (4) and (5) 
into Eq. (\ref{eq:bc}). Above, the vector $\boldsymbol{x}$ contains all the scattering coefficients.
Once these have been determined, the wavefunctions are known everywhere in the system. With these in hand, we are in a position 
to calculate both the spin-transfer torque and the MR of our system. Let us investigate the former in some detail to begin 
with. The spin-operator may be defined as
\begin{align}
\boldsymbol{S} = \psi^\dag \text{diag}(\boldsymbol{\sigma}, \boldsymbol{\sigma}^*)\psi/2,
\end{align}
and satisfies the following continuity equation in the superconducting region:
\begin{align}\label{eq:spincurrent}
&\frac{\partial \boldsymbol{S}}{\partial t} + \nabla \cdot \boldsymbol{j}_S = 0,\notag\\
\boldsymbol{j}_S &= \frac{1}{2m_S}\text{Im}\{ \Psi_S^* \nabla [\text{diag}(\boldsymbol{\sigma},\boldsymbol{\sigma}^*) \Psi_S]\}.
\end{align}
Note that there is no source term on the right hand side of the continuity equation, in contrast to the corresponding 
continuity equation in the ferromagnetic regions. The spin-current density is not conserved in the ferromagnetic regions since 
it is partially absorbed by the ferromagnetic order parameter as a spin-transfer torque. In the superconducting region, 
however, the spin-current density is conserved and is in general a tensor, since it has both a direction of flow in real 
space and a direction in spin space. When the incident spin-current from the superconducting region impinges on the interface 
to the right ferromagnet F$_\text{R}$, its transverse component must be absorbed since the spin-current deep inside the 
ferromagnetic region must be polarized parallell to its magnetization. Therefore, the spin-transfer torque density is 
equal to the components of the spin-current in the superconductor perpendicular to the $\hat{\boldsymbol{z}}$-direction, 
since $\boldsymbol{M}_\text{R} \parallel \hat{\boldsymbol{z}}$. The spin-transfer torque may be further decomposed into 
an in-plane and out-of-plane component, $\tau_\parallel$ and $\tau_\perp$, with respect to the plane with normal vector 
\begin{align}
\hat{\boldsymbol{n}}=(\ml\times\mr)/|\ml\times\mr|
\end{align}
spanned by the magnetizations. For our geometry, one infers that the in-plane component is simply the $\hat{\boldsymbol{y}}$-component 
of the spin-current in the superconductor while the out-of-plane component is the 
$\hat{\boldsymbol{x}}$-component. Therefore, we may calculate the torque in our system by first obtaining the scattering coefficients, 
constructing the spin-current of Eq. (\ref{eq:spincurrent}), and then using
\begin{align}
\tau_\parallel = \boldsymbol{j}_S\cdot \hat{\boldsymbol{y}},\; \tau_\perp = \boldsymbol{j}_S\cdot \hat{\boldsymbol{x}},\; \tau = \sqrt{\tau_\parallel^2 + \tau_\perp^2}.
\end{align}
Note that for tunnel junctions with low interface transparencies, the absorption of the transverse components of the spin-current is 
much less efficient than for metallic interfaces. In that case, the spin-current may retain components that correspond to a precession 
around the magnetization axis for a larger distance inside the ferromagnet than in high-transparency case. Here, we shall assume 
that the ferromagnetic reservoir is large enough to absorbs the transverse components fully.
\par
In order to gain insight into how the spin-transfer torque is affected by superconducting correlations, and in particular the existence 
of topological zero-energy states in the $d_{xy}$-wave case, we proceed with an approach similar to that of Ref. \cite{stiles_prb_02}. 
We consider incident electrons at the Fermi level $(\varepsilon = 0)$ from the left ferromagnet, adapting a free-electron description for 
these conduction electrons which allows us to employ Eq. (\ref{eq:wave}). In order to properly account for classical dephasing, we 
take into account the full range of transverse wavevector components. The spin-transfer torque on the right ferromagnetic region 
is then obtained by computing the spin-current in the superconductor for both incident spin-$\uparrow$ and spin-$\downarrow$ 
electrons at Fermi level, and adding their contribution with a weight-factor of 
\begin{align}
P_\sigma = (1+\sigma h/\mu_F)/2
\end{align}
meant to reflect the different magnitude of the density of states due to the finite polarization of the ferromagnet. 

The contribution to the spin-current from a given angle of incidence gives rise to an oscillatory behavior in the components of 
the spin-current which are perpendicular to the magnetization. In our case, these are the $\hat{\boldsymbol{x}}$- and
$\hat{\boldsymbol{y}}$-components since $\boldsymbol{M}_\text{R} \parallel \hat{\boldsymbol{z}}$. This oscillatory behavior 
represents a spin-precession around the $\hat{\boldsymbol{z}}$-axis upon penetration into the ferromagnetic region with a 
period of precession which is very short for transition metals (a few atomic lattice spacings) \cite{ralph_jmmm_08}. However, 
this oscillatory motion dies out due to classical dephasing which occurs when taking into account all angles of incidence 
of the spin-current. The point is that electrons reaching a given distance inside the magnet must have
travelled different path lengths to get there, leading to a destructive interference between the phases. 
\par
In addition to considering the spin-transfer torque, we also calculate the MR of the superconducting spin-valve depicted 
in Fig. \ref{fig:model}. At zero temperature, it may be defined as
\begin{align}\label{eq:MR}
\text{MR} = \frac{G_\text{P} - G_\text{AP}}{G_\text{P}},
\end{align}
where $G_\text{P/AP}$ is the conductance in the parallell/antiparallell (P/AP) magnetization configuration. Whereas the MR 
has been studied for superconducting spin-valves with conventional $s$-wave symmetries in several previous works, the case 
of a $d$-wave superconductor sandwiched between two ferromagnets has received little attention 
\cite{yoshida_prb_00, nemes_prb_08, mandal_prb_08} in comparison.
Here, we will show that the unconventional $d$-wave symmetry can induce a strongly enhanced MR effect, in particular 
when there is a formation of zero-energy surface states. Following the calculations of Refs. 
\cite{lambert_jpcm_91, dong_prb_03, yamashita_prb_03, bozovic_njp_07}, we express the normalized conductance of our 
system when a voltage $eV$ is symmetrically applied to the junction with $eV/2$ in F$_\text{L}$ and $(-eV/2)$ in $F_\text{R}$ as:
\begin{align}\label{eq:cond}
G = \sum_{\lambda\sigma} \int^{\pi/2}_{-\pi/2} \text{d}\theta \cos\theta P_\sigma \Big[ \frac{\text{Re}\{k_h^\lambda\}}{k_\text{inc}} |r_h^\lambda|^2 + \frac{\text{Re}\{k_e^\lambda\}}{k_\text{inc}} |t_e^\lambda|^2\Big].
\end{align}
where $k_\text{inc}$ is the $\hat{\boldsymbol{x}}$-component of the incident wavevector. It should be implicitly understood in 
Eq. (\ref{eq:cond}) that the contribution from incident particles with both spin-$\uparrow$ and spin-$\downarrow$ are taken 
into account, leading to different scattering coefficients $r_h^\lambda$ and $t_e^\lambda$ (the index $\sigma$ has been 
suppressed in these quantities for the sake of notation). Previous theoretical works have considered the MR in 
conventional ($s$-wave) F$\mid$S$\mid$F spin-valves for either exclusively P or AP configurations \cite{dong_prb_03, yamashita_prb_03}, 
while recently Ref. \cite{bozovic_njp_07} considered an arbitrary magnetization configuration. Here, we will investigate the 
influence of a $d$-wave superconducting order parameter on the MR, with particular emphasis on the role of zero-energy states. 
While it was shown in Ref. \cite{yoshida_prb_00} that the presence of zero-energy states in $d$-wave superconducting spin-valves 
lead to an anomalous voltage dependence of the MR effect, it remains to be clarified how the barrier transparency and the width
of the superconducting layer affect the MR. The width of the layer can be controlled accurately in the fabrication of 
experimental samples, while the interface transparency may be expected to vary from sample to sample. The spin injection
properties of heterostructures with unconventional superconductors ($p$-wave and $d$-wave) have also been studied in 
Refs. \cite{kashiwaya_prb_99, linder_prb_08, brydon_jpsj_08}.

\begin{figure*}
\centering
\resizebox{0.98\textwidth}{!}{
\includegraphics{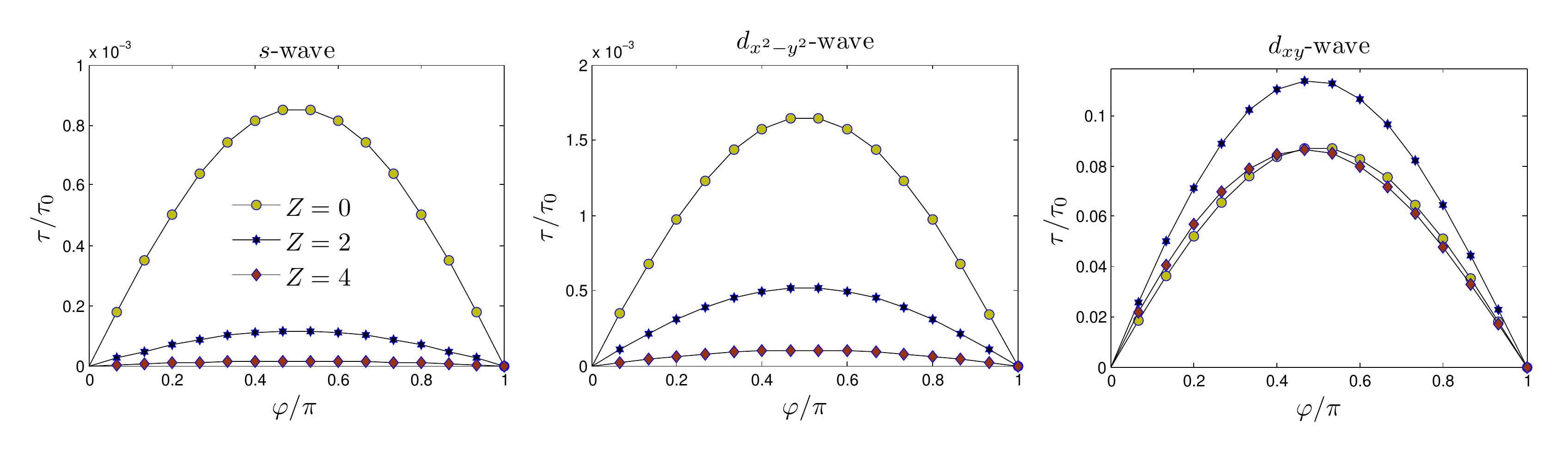}}
\caption{(Color online) Plot of the spin-transfer torque and its dependence on the misorientation angle $\varphi$. We 
give results for three different symmetries of the superconducting order parameter with interfaces characterized by 
the barrier parameter $Z$. Here, we have fixed $d_S/\xi=3.0$. }
\label{fig:torque}
\end{figure*}

\section{Results}\label{sec:results}

We now discuss our choice of parameters. The ferromagnetic regions will be modelled by taking $\mu_F = 3$ eV 
and $h = 0.9$ eV, which should be appropriate for a transition metal ferromagnet. Note that the energy-scale of this exchange field ($\sim 10^4$ K) is much larger than room-temperature ($\sim 300$ K). In the superconducting 
region, the Fermi level and gap magnitude depend on whether we have a type I $s$-wave superconductor or 
a type II $d$-wave superconductor. For a standard BCS superconductor such as Al, we may set $\Delta=1$ meV, 
$v_F=10^6$ m/s, and $\mu_S = 10$ eV. On the other hand, for a system such as a high-$T_c$ cuprate superconductor (\eg YBCO) featuring a $d$-wave order parameter, a representative choice would $\Delta=20$ meV, 
$v_F=5\times10^4$ m/s, and $\mu_S=1$ eV. We use the former set of parameter for the $s$-wave case, and 
the latter set of parameters for the $d_{xy}$- and $d_{x^2-y^2}$-cases. Moreover, we will consider the 
ballistic limit, and employ a thickness of the superconducting region equivalent to several coherence 
lengths. For the above parameters, we find that $\xi \simeq  1$ nm in the $d$-wave case, which indicates 
that the thickness of the superconducting layer should be of several nm, which is experimentally 
feasible. The electron mass is taken to be close to its bare value in the F regions, $m_F=0.5$ MeV, while in 
the superconductor it is calculated according to the free-electron dispersion $m_S = 2\mu_S/r_v^2$, where 
$r_v=1/v_F$ is the ratio of the velocity of light and the Fermi velocity (note that we set $c=1$).
\par
In order to justify a non-selfconsistent approach (see also discussion in Sec. \ref{sec:discussion}), we 
consider for the spin-transfer torque a rather thick layer $d_S/\xi=3.0$ and consider several values for 
the barrier transparency $Z$. Also, we focus only on small bias voltages $eV/\Delta_0\ll1$ when 
calculating the MR, since larger voltages $eV\sim\Delta_0$ lead to a substantial spin accumulation in 
the superconducting region which strongly suppresses the gap and finally destroys superconductivity 
itself \cite{takahashi_prl_99}. For $eV/\Delta_0\ll1$, a self-consistent solution of the gap should 
not be required \cite{yoshida_prb_00}.

\subsection{Spin-transfer torque}
We now proceed to investigate the behavior of the spin-transfer torque in the spin-valve setup. The normalization 
constant 
\begin{align}
\tau_0 = \sqrt{2m_F\mu_F}/m_S
\end{align}
is introduced below. First, we plot the magnitude $\tau$ of the spin-transfer torque versus the misorientation angle 
$\varphi$ in Fig. \ref{fig:torque}. The results are shown for several order parameter symmetries in the superconducting 
region. The $s$-wave and $d_{x^2-y^2}$-wave symmetries do not feature zero-energy states whereas the $d_{xy}$-wave 
symmetry does. From Fig. \ref{fig:torque}, one infers two important pieces of information. Firstly, it is seen that 
the dependence on the misorientation angle essentially follows the usual sinusoidal behavior, regardless of the 
symmetry of the superconducting order parameter. Secondly, whereas the magnitude of the spin-transfer torque rapidly 
decreases with increasing barrier strength $Z$ in the $s$-wave and $d_{x^2-y^2}$-wave cases, the torque is strongly 
enhanced and more resilient towards an increase of $Z$ in the $d_{xy}$-wave case. In fact, in this case the spin-transfer 
torque shows a weak maximum as a function of $Z$ before it slowly decreases for large values of $Z$. This is a direct 
consequence of the existence of zero-energy surface states, as will be discussed below. 
\par
\begin{figure}[b!]
\centering
\resizebox{0.5\textwidth}{!}{
\includegraphics{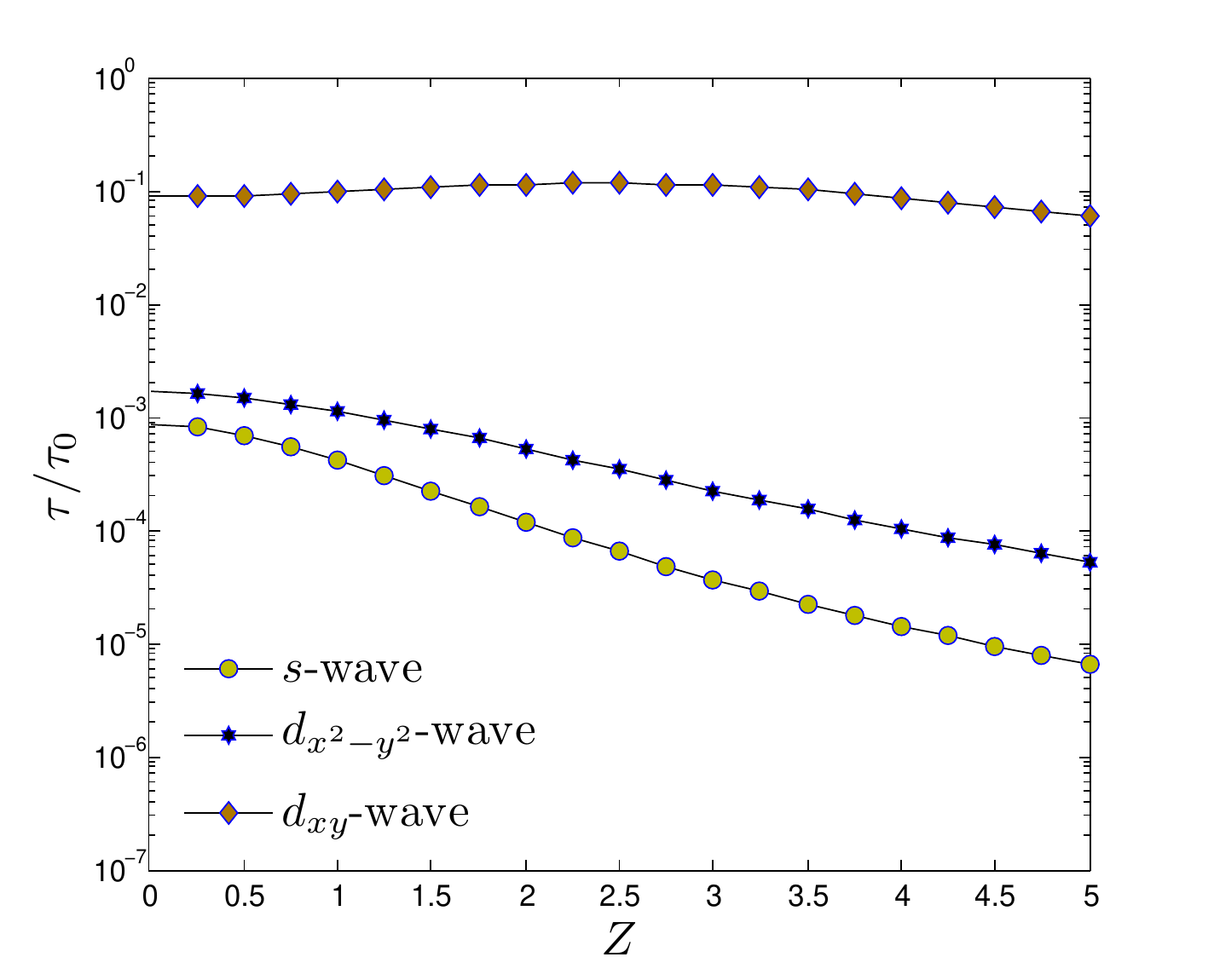}}
\caption{(Color online) Plot of the spin-transfer torque and its dependence on the interface transparency $Z$. 
Here, we have set $d_S/\xi=3.0$ and $\varphi=\pi/2$. }
\label{fig:torque_Z}
\end{figure}

To investigate this matter further, we plot in Fig. \ref{fig:torque_Z} the spin-transfer torque as a function of $Z$ 
with a fixed misorientation angle $\phi=\pi/2$. Since the magnitude of the torque differs greatly for the various 
symmetries, we use a logarithmic scale for the ordinate axis. As seen, the torque has a smooth peak at $Z\simeq2$ for 
the $d_{xy}$-wave symmetry. Also, its magnitude is considerably larger -- the ratio of the MR in the $d_{xy}$-wave 
and $s$-wave case is $\sim 10^3$ for $Z\simeq2$. In contrast, the torque decays monotonically with $Z$ for all values 
of $Z > 0$ in the $s$-wave and $d_{x^2-y^2}$-wave cases. These symmetries do not host zero-energy states at the 
interfaces, and thus an increase in barrier strength simply amounts to a reduction of the current since tunneling 
becomes prohibited, and concomitantly the magnitude of the torque decreases. However, when zero-energy states are 
present the quasiparticle current through the superconductor is strongly enhanced at the Fermi level compared to 
\eg the fully gapped $s$-wave case, and the resulting spin-transfer torque increases in magnitude. This explains 
the monotonic decay of the spin-transfer torque in the $s$- and $d_{x^2-y^2}$-cases when $Z$ is increased, as 
well as the large difference in overall amplitude of the spin-transfer torque between, on the one hand the  
$s$- and $d_{x^2-y^2}$-cases, and on the other hand the $d_{xy}$-case. The nonmonotonic variation of 
$\tau/\tau_0$ as a function of $Z$ for the $d_{xy}$-case is a more subtle matter, to which we will return in Sec.
\ref{sec:discussion}.  

\begin{figure}[b!]
\centering
\resizebox{0.4\textwidth}{!}{
\includegraphics{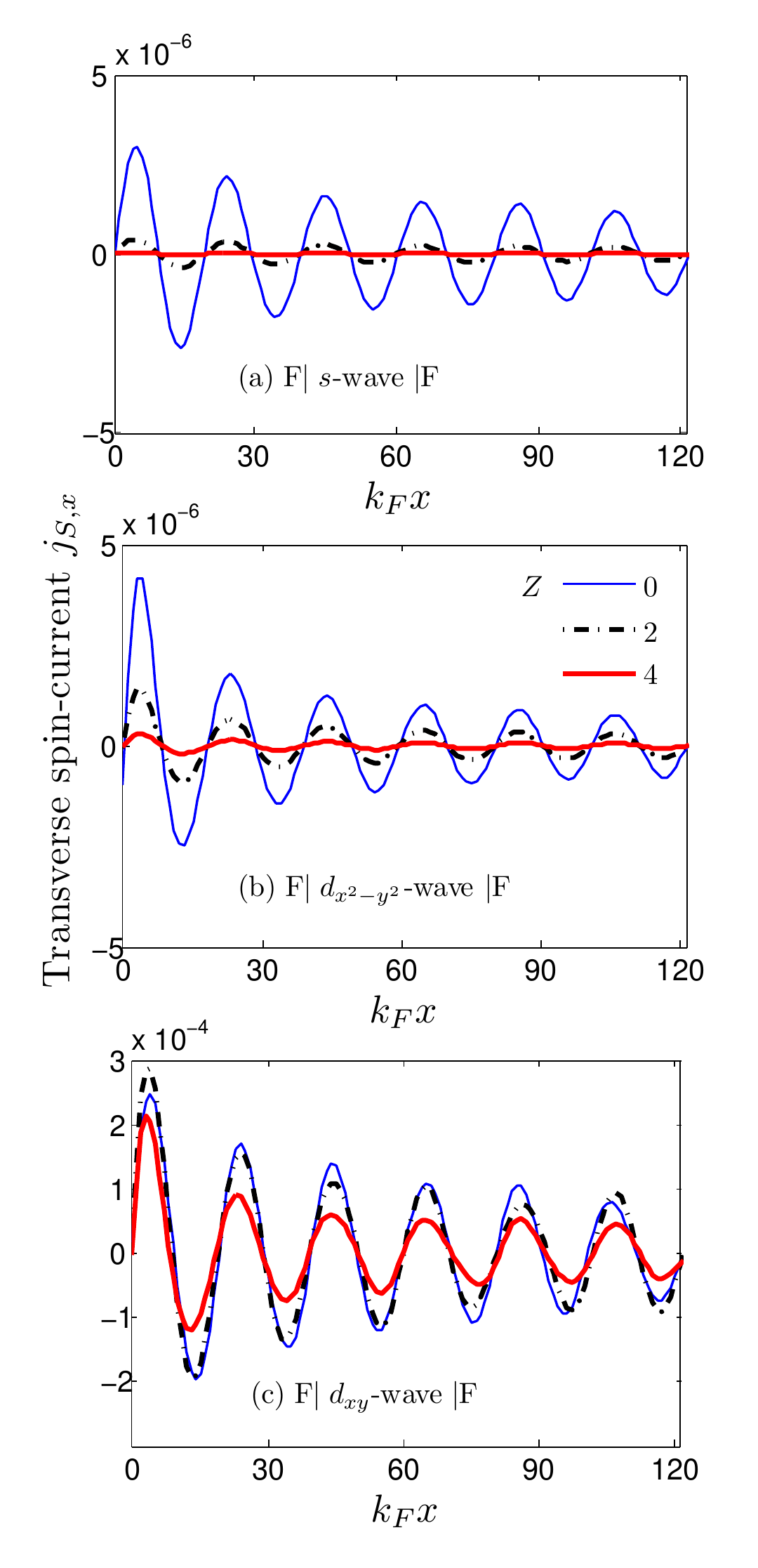}}
\caption{(Color online) Plot of the decay of the transverse ($\hat{\boldsymbol{x}}$-component) part of the spin-current 
in the right ferromagnetic region for several values of the interface transparency $Z$. Here, we have set $d_S/\xi=3.0$ 
and $\varphi=\pi/2$. }
\label{fig:decay}
\end{figure}

Let us also address the decay of the spin-current inside the ferromagnetic region. As previously mentioned, the 
transverse components of the spin-current will in general precess around the magnetic order parameter, but decay 
rapidly once inside the ferromagnetic region. In Fig. \ref{fig:decay}, we investigate how this decay depends on 
the barrier parameter $Z$ for various order parameter symmetries in the superconducting regions. We only give the 
results for the $\hat{\boldsymbol{x}}$-component of the spin-current, as the result for the 
$\hat{\boldsymbol{y}}$-component is basically the same. Note that the $\hat{\boldsymbol{z}}$-component of the 
spin-current remains constant in the ferromagnetic region. As seen, the presence of superconducting correlations 
lead to no qualitatively new behavior with respect to the spatial decay of the transverse components of the spin 
current. The spin-current oscillates rapidly on a scale comparable to several Fermi wavelengths $1/k_F$ where we 
have defined $k_F = \sqrt{2m_F\mu_F}$. In terms of magnitude, it is seen that the spin-current is strongly 
enhanced in the $d_{xy}$-wave case compared to the  $s$-wave and $d_{x^2-y^2}$-wave symmetry.

\subsection{Magnetoresistance}\label{sec:MR}

\begin{figure*}
\centering
\resizebox{0.98\textwidth}{!}{
\includegraphics{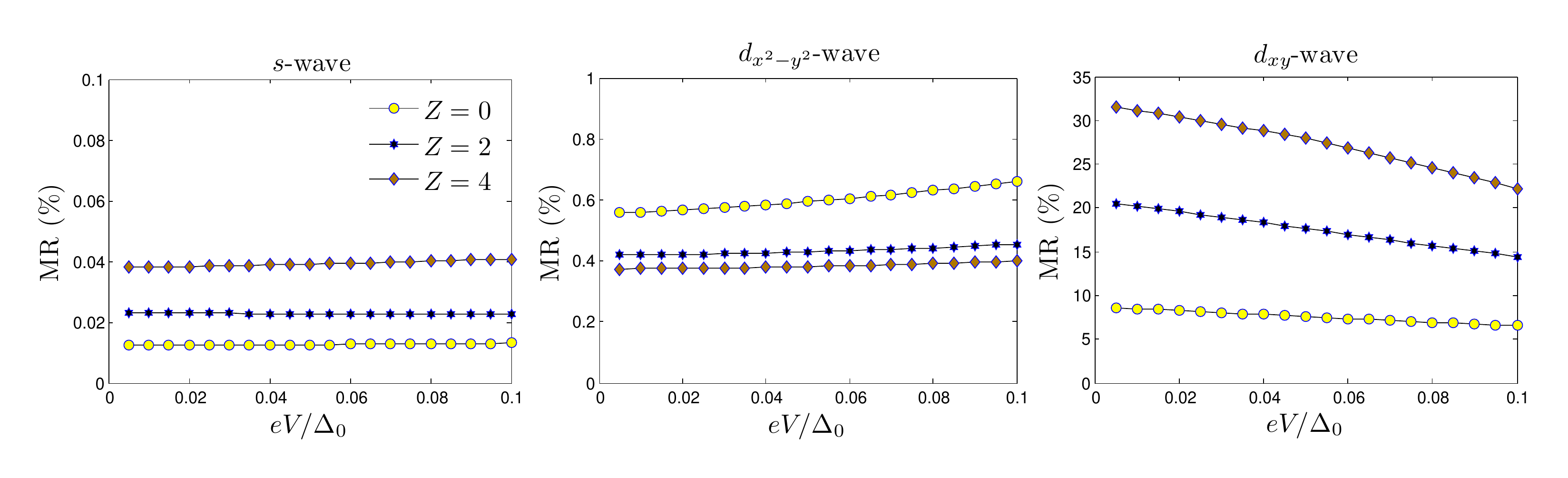}}
\caption{(Color online) Plot of the MR for several choices of symmetry for the superconducting order parameter and 
several barrier transparencies $Z$. Here, we have set $d_S/\xi=3.0$. }
\label{fig:MR}
\end{figure*}

Next, we consider the MR effect in our superconducting spin-valve setup, defined by Eq. (\ref{eq:MR}). In Fig. \ref{fig:MR}, 
we plot the MR in percentage versus applied bias voltage in the regime $eV\ll \Delta_0$ where one may disregard spin 
accumulation. Similarly to the spin-transfer torque, the $d_{xy}$-wave case behaves qualitatively differently from 
the $s$-wave and $d_{x^2-y^2}$-wave symmetries. In the former case, the MR decreases in magnitude upon increasing the 
bias voltage, while 
in the latter case the MR increases slowly. This may again be understood by the enhancement of the quasiparticle current 
at $eV =0$ due to the presence of zero-energy states. When $eV \to 0$, it is seen that the MR is enhanced by a factor
$\sim 10^3$ in the $d_{xy}$-wave case compared to the fully gapped $s$-wave case, similarly to the increase of 
spin-transfer torque discussed in the previous section.

\begin{figure}[b!]
\centering
\resizebox{0.5\textwidth}{!}{
\includegraphics{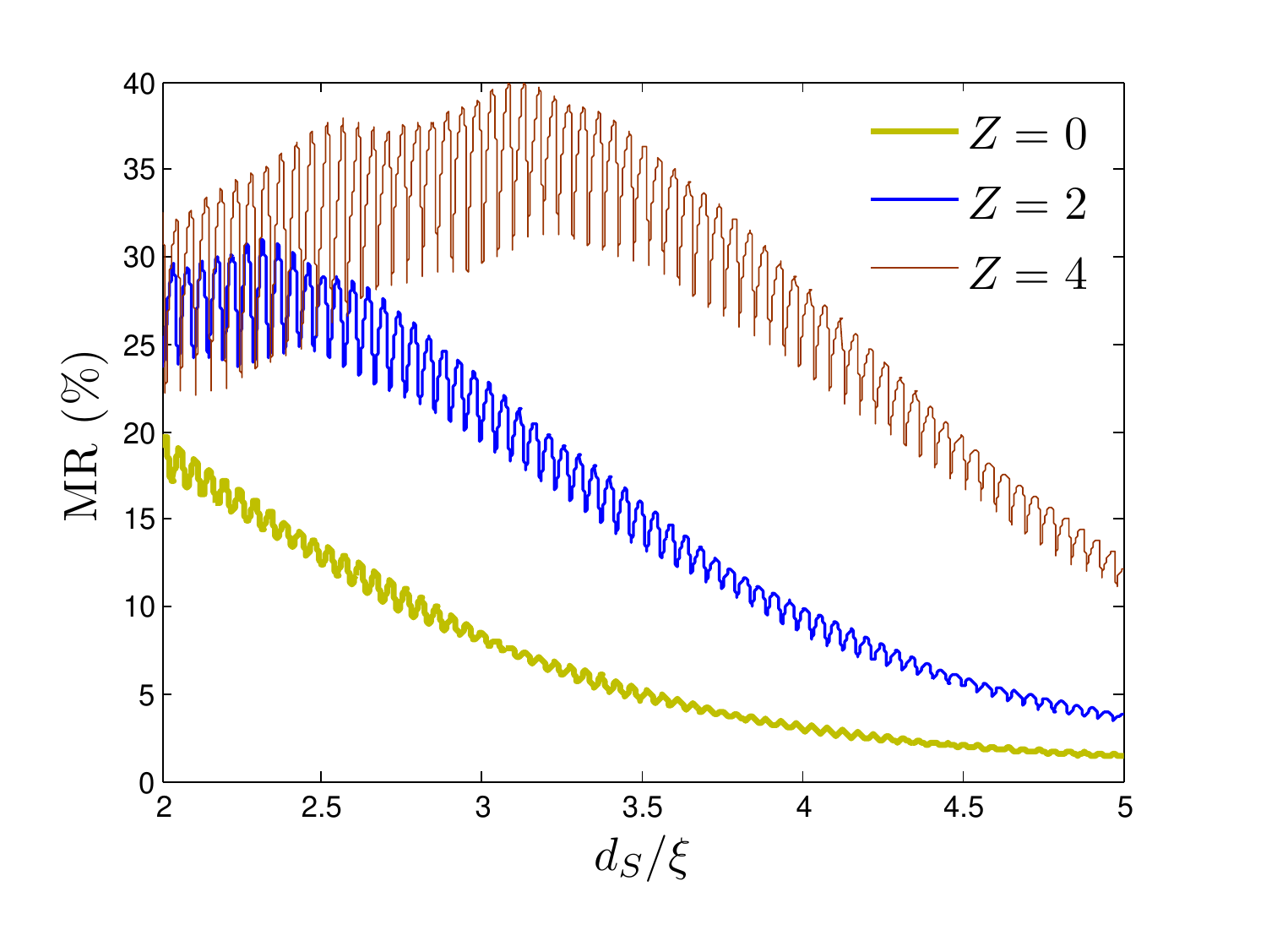}}
\caption{(Color online) Plot of the MR for the $d_{xy}$-wave symmetry with $eV/\Delta_0 = 0.01$.}
\label{fig:MR_d}
\end{figure}

Interestingly, we find that the MR shows a non-monotonic behavior upon changing the interlayer width $d_S$ of the 
superconducting region in the $d_{xy}$-wave case. This is shown in Fig. \ref{fig:MR_d}, where we plot the MR as a 
function of $d_S/\xi$ for $eV/\Delta=0.01$. In the case $Z=0$ (perfect barrier transparency), the MR decays just 
like in the $s$-wave case, only with a much enhanced magnitude. However, for $Z\neq0$ the MR undergoes a smooth 
maximum at a finite value of $d_S/\xi$, before the decaying behavior sets in. 

We propose the following explanation for this observation. The topological zero-energy surface states are bound 
to the surface, and penetrate a distance comparable to $\xi$ into the superconducting region. Therefore, they cannot 
contribute to the transport properties of spin-valves with widths $d_S/(2\xi) \gg 1$. This is the reason for the 
decay which sets in at a finite value of $d_S$ in Fig. \ref{fig:MR_d}, signalling precisely the loss of contribution 
to transport from such zero-energy states. It is also seen that the MR displays rapid oscillations as a function 
of $d_S$, pertaining to the formation of resonant transmission states in the superconducting region. We have 
checked that upon increasing the voltage $eV$, the value of $d_S/\xi$ at which the MR peaks is shifted towards 
smaller values, which is consistent with the fact that the influence of the zero-energy states gradually 
vanishes when moving away from the Fermi level.
\par
Note that the expression for the conductance in Eq. (\ref{eq:cond}) applies to a situation where a voltage 
is symmetrically applied to the F$\mid$S$\mid$F junction, with incoming electrons/holes from the left/right 
side, and corresponds to a situation where a current is driven through the entire system. This is distinct 
from a situation where a positive bias $eV$ is applied to one of the F leads while the superconductor and 
the other lead are both grounded. In the present case, we are interested in how the spin-transfer torque and 
MR are altered depending on the magnetization configuration and pairing symmetry in the superconductor. This 
is probed simply by passing a current through the F$\mid$S$\mid$F junction. If the main aim was to investigate 
crossed Andreev reflection signals, it would be more beneficial to study the aforementioned situation where 
a voltage is applied only between one of the leads and the superconductor and then observe a possible 
current-response in the other lead. Such an effect would be indicative of non-local scattering processes 
such as crossed Andreev reflection.

\section{Discussion}\label{sec:discussion}

Throughout this study, we have made some simplifying assumptions in order to proceed with a partially analytical approach. For instance, 
we have adopted the standard step-function approximation of the superconducting order parameter $\Delta_0$ at the F$\mid$S interfaces, 
which should be a reasonable approximation as long as $d_S\geq2\xi$. A full numerical approach would most likely reveal that $\Delta_0$ 
is subject to a partial depletion near the interfaces, but we do not expect that this will alter our results qualitatively. Also, we 
have disregarded any thermal effects and their influence on the spin-transfer torque since we consider the zero-temperature limit. The 
presence of a superconductor demands that $T<T_c$ where $T_c$ is the superconducting critical temperature, and it is thus reasonable 
that thermal fluctuations should be neglible in the regime $T \ll T_c$. In conventional experiments at room-temperature with F$\mid$N$\mid$F 
spin-valves, thermal fluctuations play an important role in particular close to the P or AP alignment $\varphi=\{0,\pi\}$ since no 
spin-transfer torque is possible for perfect P or AP alignment. However, by considering fluctuations around a magnetization axis, 
a spin-transfer torque becomes applicable. Typically, the magnetization in one of the F layers is considered to be static with a 
high magnetization density and correspondingly weak fluctuations, while the other one is dynamic and considerably influenced by 
the thermal fluctuations \cite{skadsem_privatecommunication}.
\par
In Ref. \cite{giazotto_prl_06}, it was shown that in the idealized situation of fully polarized ferromagnets (i.e. half-metals) and 
perfect interfaces ($Z=0$), the crossed Andreev reflection (CAR) process could lead to very large MR values in an F$\mid$S$\mid$F 
spin-valve when the superconductor had a conventional $s$-wave symmetry. When the superconductor has a $d_{xy}$-wave symmetry, we 
find that at normal incidence $\theta=0$ the probability for a CAR process is exactly zero independent of $\{Z,d_S,eV\}$ and 
regardless of whether the magnetization configuration is in the P or AP alignment. In the $d_{x^2-y^2}$-wave case, however, 
the probability for CAR scattering remains finite at normal incidence. This finding is in agreement with the results reported 
very recently in Ref. \cite{herrera_prb_09}, namely that elastic-cotunneling (EC) processes are favored along the nodal lines 
of the superconducting order parameter while CAR processes is enhanced along directions where the magnitude of the gap is maximal. 
Since $\theta=0$ corresponds to the nodal line of a $d_{xy}$-wave order parameter, the vanishing CAR amplitude is consistent 
with the above statement.
\par
\begin{figure}[t!]
\centering
\resizebox{0.5\textwidth}{!}{
\includegraphics{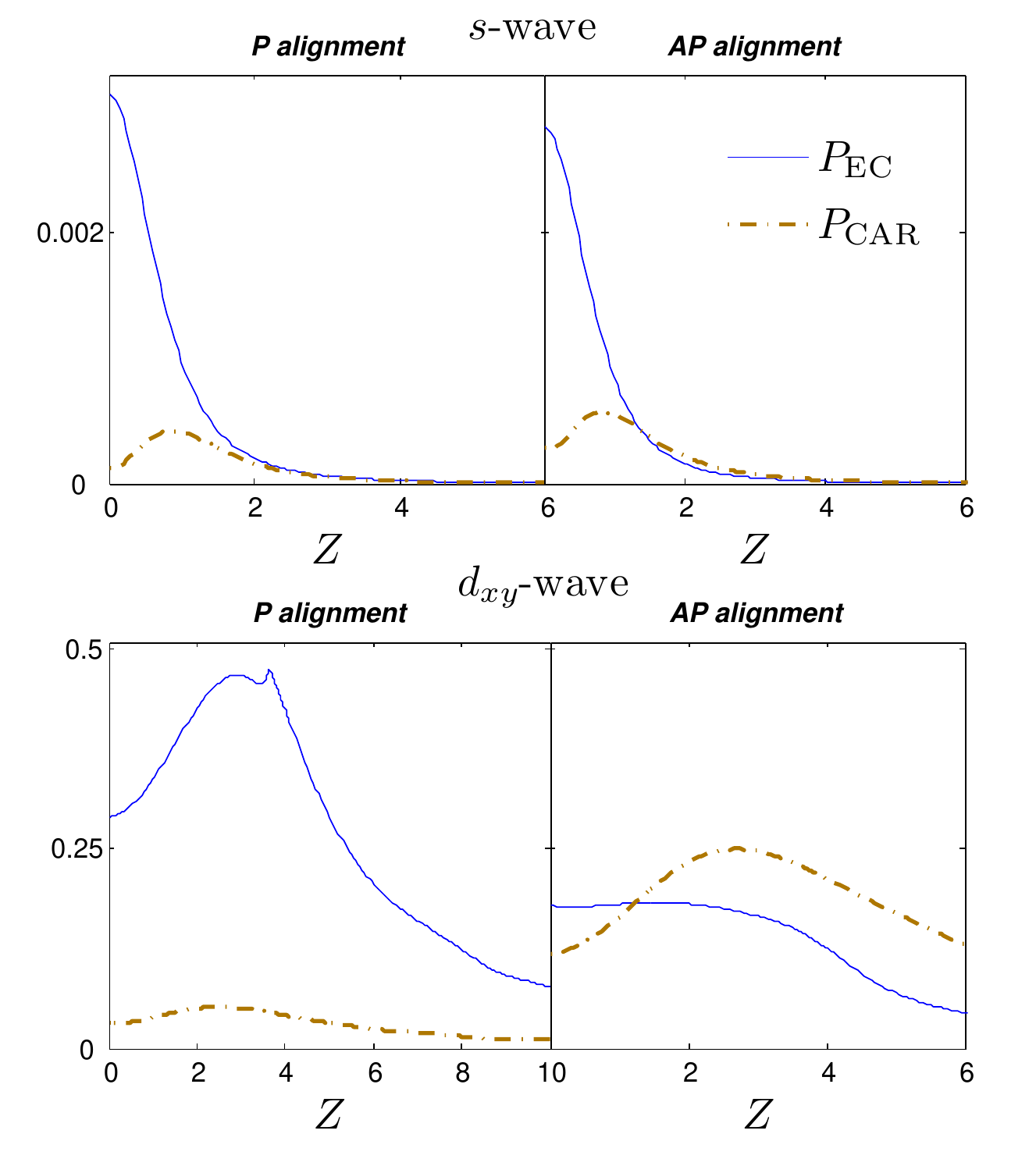}}
\caption{(Color online) Plot of the probability measures for EC and CAR processes as a function of the barrier parameter 
$Z$, with fixed $eV/\Delta_0=0.01$ and $d_S/\xi=3.0$.}
\label{fig:Prob}
\end{figure}

The absence of CAR at normal incidence for the $d_{xy}$-wave symmetry suggests that the large MR effect is mainly attributed 
to quasiparticle transport through the superconductor facilitated by zero-energy states \cite{Tanaka}. However, in 
order to draw a firm conclusion about this, we must take into account all angles of incidence $\theta$. To investigate this 
quantitatively, let us introduce some helpful quantities. Consider now only incoming spin-$\uparrow$ electrons, which dominate 
the transport properties due to the polarization factors $P_\sigma$ and the strong exchange field $h/\mu_F \simeq 0.3$. For 
a given angle of incidence $\theta$, the probability of an elastic co-tunneling process (EC) and a CAR process are given by 
\begin{align}
P_\text{EC}(\theta) = \frac{\text{Re}\{k_e\}}{k_\text{inc}} |t_e|^2,\; P_\text{CAR}(\theta) = \frac{\text{Re}\{k_h\}}{k_\text{inc}} |t_h|^2,
\end{align}
respectively. As a total measure of the probability for an EC and CAR process, we introduce
\begin{align}
P_\mathrm{J} = \frac{1}{N_\theta}\sum_{\theta=0}^{\pi/2} P_\mathrm{J}(\theta),\; j=\{\mathrm{EC}, \mathrm{CAR}\},
\end{align}
where the summation is over all incoming angles and $N_\theta$ is the number of angles summed over (we set $N_\theta=157$ in 
what follows). In this way, $P_\mathrm{J} \in [0,1]$ is a probability measure for the EC and CAR processes when taking into 
account all angles of incidence. We plot $P_\mathrm{J}$ for $eV/\Delta_0=0.01$ and a junction of width $d_S/\xi=3.0$ in 
Fig. \ref{fig:Prob} in order to understand better how they influence the MR effect of the system. We have checked that 
these quantities remain virtually constant in the low energy regime $eV/\Delta_0 \ll 1$. Consider first the $s$-wave case. 
As seen, the EC process dominates for small $Z$, regardless of whether the magnetization configuration is P or AP. Increasing 
$Z$ (corresponding to tunnel junction), the EC and CAR processes become essentially equal in magnitude. Since there is so 
little difference quantitatively between the P and AP configuration one expects a small MR effect for the $s$-wave case, 
which is consistent with our results presented in Sec. \ref{sec:MR}.

Consider now the $d_{xy}$-wave case. In the P alignment, the EC process again dominates the CAR process, but in contrast to the 
$s$-wave case this now occurs for all $Z$. The situation changes drastically upon reversing the alignment to AP, where the 
contribution from the CAR process becomes dominant at large values of $Z$. The most important point illustrated in Fig. \ref{fig:Prob} 
for the $d_{xy}$-wave case is that the EC (CAR) process is strongly
enhanced in the P (AP) alignment compared to the AP (P) case. The EC process is facilitated in this context precisely due to the 
existence of zero-energy states which exist near the interface and extend a distance of order $\sim \xi$ from each interface 
into the bulk of the superconductor. Simultaneously, the CAR process is strongly suppressed in the P alignment because of the 
strong exchange field that makes spin-$\downarrow$ holes less available in the right ferromagnet. In the AP alignment, however, 
the CAR process can be enhanced by the existence of zero-energy states since there are more spin-$\downarrow$ holes available, 
with a concomitant reduction of the EC process. The conclusion is that whereas the EC and CAR processes display a weak 
sensitivity to the magnetization configuration in the $s$-wave case, the presence of zero-energy states in the $d_{xy}$-wave 
case strongly enhances the EC and CAR processes. This enhancement is not present in the $s$-wave junction. Upon combining the 
enhancement of EC case with the suppression of CAR in the P alignment of the $d_{xy}$-wave, we obtain a very large MR effect. 
Note that it is not the dominance of EC over CAR itself that leads to the large MR effect -- it is the difference between 
these contributions when comparing the P and AP alignments.
\par
Finally, note that although the CAR process may dominate over the EC process in a certain parameter regime (\eg for large $Z$ 
in the AP alignment of Fig. \ref{fig:Prob}), the conductance of the junction is always positive. The CAR and EC processes do 
contribute to the conductance with opposite signs, as may be seen by expressing the conductance in Eq. (\ref{eq:cond}) with 
the CAR and normal reflection amplitudes via the conservation of probability currents, but the conductance still remains 
positive in our case of a symmetrically applied voltage to the junction. This follows from simple current-conservation 
upon applying a bias voltage difference between the leads. In a scenario where a voltage is applied only to, say, the 
left lead while the superconductor and the right lead are grounded, the cross-conductance may take any sign depending 
on whether CAR or EC dominates the non-local transport properties.\cite{herrera_prb_09}
\par
The results of Fig. \ref{fig:Prob} are also relevant to the non-monotonic behavior of the spin-torque transfer
as a function of $Z$ found in Fig. \ref{fig:torque_Z}, as we discuss below. The EC process contributes positively to the magnitude 
of the current, hence also to the torque, and is seen to be maximal at a finite value of $Z$ in the $d_{xy}$-wave case.

 We propose the following explanation for the reason why $P_{\rm{EC}}(Z)$ is non-monotonic in the $d_{xy}$-case for parallell alignment, while it is monotonically decaying in the other cases: When there are no ZES, the EC process cannot be facilitated by any subgap states in the superconducting region and it simply decays with increasing $Z$. Consider now the case with ZES, which facilitates the EC process. In the AP alignment, the EC process is hindered by the density of states mismatch in the two ferromagnets as compared to the P alignment, where there is no Fermi-vector mismatch associated with tunneling of a spin-$\sigma$ electron from the left ferromagnet to the right ferromagnet via the superconductor. In light of this, it is reasonable to expect that the ideal circumstances for the EC process is the P alignment in the presence of ZES, which is seen to be the case from Fig. \ref{fig:Prob}. Note that the gap nodes coincide with normal incidence in the $d_{xy}$-wave case, while the gap amplitude is maximal at normal incidence in the $d_{x^2-y^2}$-wave case, which also contributes to the suppression of EC in the latter case. The reason for why we only see a non-monotonous behavior of $P_\text{EC}$ in the P alignment for the $d_{xy}$-wave case is thus that it is only in this case that the EC process is actually favorable without any suppressing mechanisms except for the barrier strength $Z$. In all the other cases, there are additional mechanisms which hinder the EC process, which evidently leads to a pure decay of the probability for this scattering process. Although the above discussion has been for the P case, it bears upon the non-collinear situation $\varphi \neq \{0,\pi\}$. The reason for this is that such a scenario may be considered as a superposition of the P and AP alignments, such that the non-monotonous behavior should be seen for any $\varphi$ which is not too close to $\pi$. The exact microscopic mechanisms behind the non-monotonous behavior of the EC process are very difficult to extract since our approach is partly numerical, but we believe the general arguments given above should provide a basic understanding of the situation.

\section{Summary}\label{sec:summary}

To summarize, we have investigated the spin-transfer torque and the magnetoresistance of a ballistic ferromagnet$\mid$superconductor$\mid$ferromagnet spin-valve, allowing for an arbitrary magnetization 
misorientation and also including the possibility of a $d$-wave symmetry for the superconducting order 
parameter. Our approach accounts properly for both Andreev reflection and proximity-induced triplet 
correlations that are generated when 
the magnetizations are non-collinear. Our main finding is that the torque and magnetoresistance are both 
strongly enhanced when topological zero-energy states are present at the interfaces, which is relevant for 
the $d_{xy}$-wave case. Also, we have found a strong oscillatory and non-monotonous behavior of the 
magnetoresistance as a function of the width $d_S$ of the superconductor. The reason for this is a 
gradual vanishing of the contribution to transport from zero-energy states as the layer size 
increases from $d_S\sim 2\xi$ to $d_S\gg 2\xi$, where $\xi$ is the superconducting coherence length.

\section*{Acknowledgments}
We thank A. Brataas, H. J. Skadsem, and H. Haugen for helpful discussions. J.L. and A.S. were supported 
by the Norwegian Research Council Grant Nos. 158518/431, 158547/431, (NANOMAT), 
and 167498/V30 (STORFORSK).
T.Y. acknowledges support by JSPS.

\appendix


\begin{thebibliography}{99}

\bibitem{zutic_rmp_04} I. Zutic, J. Fabian, and S. Das Sarma, Rev. Mod. Phys. \textbf{76}, 323 (2004).

\bibitem{tserkovnyak_rmp_05} Y. Tserkovnyak, A. Brataas, G. E. Bauer, and B. I. Halperin, Rev. Mod. Phys. \textbf{77}, 1375 (2005).

\bibitem{brataas_physrep_06} A. Brataas, G. E.W. Bauer, and P. J. Kelly, Phys. Rep. \textbf{427}, 157 (2006).

\bibitem{Maekawa} S. Maekawa (Ed.), Concept in Spin Electronics, Oxford University Press, Oxford, 2006.

\bibitem{Tatara} G. Tatara, H. Kohno, and J. Shibata, Phys. Rep. \textbf{468}, 213 (2008).

\bibitem{ralph_jmmm_08} D. C. Ralph and M. D. Stiles, J. Magn. Magn. Mat. \textbf{320}, 1190 (2008).

\bibitem{sankey_nphys_08} J. C. Sankey, Y.-T. Cui, J. Z. Sun, J. C. Slonczewski, R. A. Buhrman, D. C. Ralph, Nature Physics \textbf{4}, 67 (2008).

\bibitem{slonczewski_jmmm_96} J. Slonczewski, J. Magn. Magn. Mater. \textbf{150}, 13 (1995).

\bibitem{myers_science_99} E. B. Myers, D. C. Ralph, J. A. Katine, R. N. Louie, and R. A. Buhrman, Science \textbf{285}, 687 (1999).

\bibitem{katine_prl_00} J. A. Katine, F. J. Albert, R. A. Buhrman, E. B. Myers, and D. C. Ralph, Phys. Rev. Lett. \textbf{84}, 3129 (2000).

\bibitem{tserkovnyak_prb_02} Y. Tserkovnyak and A. Brataas, Phys. Rev. B \textbf{65}, 094517 (2002).

\bibitem{waintal_prb_01} X. Waintal and P. W. Brouwer, Phys. Rev. B \textbf{63}, 220407 (2001).

\bibitem{waintal_prb_02} X. Waintal and P. W. Brouwer, Phys. Rev. B \textbf{65}, 054407 (2002).

\bibitem{lofwander_prl_05} T. L{\"o}fwander, T. Champel, J. Durst, and M. Eschrig, Phys. Rev. Lett. \textbf{95}, 187003 (2005).

\bibitem{Zhao} E. Zhao and J. A. Sauls, Phys. Rev. B \textbf{78}, 174511 (2008).
\bibitem{bergeretrmp} F. S. Bergeret, A. F. Volkov, and K. B. Efetov, 
Rev. Mod. Phys. \textbf{77}, 1321 (2005).

\bibitem{buzdinrmp} A. I. Buzdin, Rev. Mod. Phys. \textbf{77}, 935 (2005).

\bibitem{Braude} V. Braude and Yu. V. Nazarov, Phys. Rev. Lett. \textbf{98}, 077003 (2007).

\bibitem{Houzet} M. Houzet and A. I. Buzdin, Phys. Rev. B \textbf{76}, 060504(R) (2007).

\bibitem{Hu} C.-R. Hu, Phys. Rev. Lett. \textbf{72}, 1526 (1994).

\bibitem{Tanaka} Y. Tanaka and S. Kashiwaya, Phys. Rev. Lett. \textbf{74}, 3451 (1995).

\bibitem{Ryu} S. Ryu and Y. Hatsugai, Phys. Rev. Lett. \textbf{89}, 077002 (2002).

\bibitem{btk} G. E. Blonder, M. Tinkham, and T. M. Klapwijk, Phys. Rev. B \textbf{25}, 4515 (1982).

\bibitem{linder_prb_07} J. Linder and A. Sudb{\o}, Phys. Rev. B \textbf{75}, 134509 (2007).

\bibitem{xiao_prb_08} J. Xiao, G. E. W. Bauer, and A. Brataas, Phys. Rev. B \textbf{77}, 224419 (2008).

\bibitem{eschrig_naturephysics_08} M. Eschrig and T. L{\"o}fwander, Nat. Phys. \textbf{4}, 138 (2008).

\bibitem{Brataas} A. Brataas, Yu. V. Nazarov, and G. E. W. Bauer, Phys. Rev. Lett. {\bf 84}, 2481 (2000); 
Eur. Phys. J. B \textbf{22}, 99 (2001).

\bibitem{stiles_prb_02} M. D. Stiles and A. Zangwill, Phys. Rev. B \textbf{66}, 014407 (2002).

\bibitem{yoshida_prb_00} N. Yoshida, Y. Tanaka, J. Inoue, and S. Kashiwaya, Phys. Rev. B \textbf{63}, 024509 (2000).

\bibitem{nemes_prb_08} N. M. Nemes, M. Garca-Hernndez, S. G. E. te Velthuis, A. Hoffmann, C. Visani, 
J. Garcia-Barriocanal, V. Pea, D. Arias, Z. Sefrioui, C. Leon, and J. Santamara, Phys. Rev. B \textbf{78}, 094515 (2008).

\bibitem{mandal_prb_08} S. Mandal, R. C. Budhani, J. He, and Y. Zhu, Phys. Rev. B \textbf{78}, 094502 (2008).


\bibitem{lambert_jpcm_91} C. J. Lambert, J. Phys.: Condens. Matter, \textbf{3}, 6579 (1991).

\bibitem{dong_prb_03} Z. C. Dong, R. Shen, Z. M. Zheng, D. Y. Xing, and Z. D. Wang, Phys. Rev. B \textbf{67}, 134515 (2003).

\bibitem{yamashita_prb_03} T. Yamashita, H. Imamura, S. Takahashi, and S. Maekawa, Phys. Rev. B \textbf{67}, 094515 (2003).

\bibitem{bozovic_njp_07} M. Bozovic and Z. Radovic, New Jour. Phys. \textbf{9}, 264 (2007).


\bibitem{kashiwaya_prb_99} S. Kashiwaya, Y. Tanaka, N. Yoshida, and M. R. Beasley, 
Phys. Rev. B \textbf{60}, 3572 (1999).

\bibitem{linder_prb_08} J. Linder, T. Yokoyama, Y. Tanaka, and A. Sudb{\o}, 
Phys. Rev. B \textbf{78}, 014516 (2008).

\bibitem{brydon_jpsj_08} P. M. R. Brydon, D. Manske, M. Sigrist, 
J. Phys. Soc. Jpn. \textbf{77}, 103714 (2008).

\bibitem{takahashi_prl_99} S. Takahashi, H. Imamura, and S. Maekawa, Phys. Rev. Lett. \textbf{82}, 3911 (1999).

\bibitem{giazotto_prl_06} F. Giazotto, F. Taddei, F. Beltram, and R. Fazio, Phys. Rev. Lett. \textbf{97}, 087001 (2006).

\bibitem{herrera_prb_09} W. J. Herrera, A. Levy Yeyati, and A. Martin-Rodero, Phys. Rev. B \textbf{79}, 014520 (2009).

\bibitem{skadsem_privatecommunication} H. J. Skadsem, private communication (2009).

\end{thebibliography}
\end{document}